\documentstyle[12pt]{article}
\begin{document}
\thispagestyle{empty}
\newcommand{\be}{\begin{equation}}
\newcommand{\ee}{\end{equation}}
\newcommand{\sect}[1]{\setcounter{equation}{0}\section{#1}}
\newcommand{\vs}[1]{\rule[- #1 mm]{0mm}{#1 mm}}
\newcommand{\hs}[1]{\hspace{#1mm}}
\newcommand{\mb}[1]{\hs{5}\mbox{#1}\hs{5}}
\newcommand{\bea}{\begin{eqnarray}}
\newcommand{\eea}{\end{eqnarray}}
\newcommand{\wt}[1]{\widetilde{#1}}
\newcommand{\ux}[1]{\underline{#1}}
\newcommand{\ov}[1]{\overline{#1}}
\newcommand{\sm}[2]{\frac{\mbox{\footnotesize #1}\vs{-2}}
           {\vs{-2}\mbox{\footnotesize #2}}}
\newcommand{\prt}{\partial}
\newcommand{\eps}{\epsilon}\newcommand{\p}[1]{(\ref{#1})}
\newcommand{\R}{\mbox{\rule{0.2mm}{2.8mm}\hspace{-1.5mm} R}}
\newcommand{\Z}{Z\hspace{-2mm}Z}
\newcommand{\cd}{{\cal D}}
\newcommand{\cg}{{\cal G}}
\newcommand{\ck}{{\cal K}}
\newcommand{\cw}{{\cal W}}
\newcommand{\vj}{\vec{J}}
\newcommand{\vl}{\vec{\lambda}}
\newcommand{\vz}{\vec{\sigma}}
\newcommand{\vt}{\vec{\tau}}
\newcommand{\poiss}{\stackrel{\otimes}{,}}
\newcommand{\tx}{\theta_{12}}
\newcommand{\tb}{\overline{\theta}_{12}}
\newcommand{\zw}{{1\over z_{12}}}
\newcommand{\sqp}{{(1 + i\sqrt{3})\over 2}}
\newcommand{\sqm}{{(1 - i\sqrt{3})\over 2}}
\newcommand{\NP}[1]{Nucl.\ Phys.\ {\bf #1}}
\newcommand{\PLB}[1]{Phys.\ Lett.\ {B \bf #1}}
\newcommand{\PLA}[1]{Phys.\ Lett.\ {A \bf #1}}
\newcommand{\NC}[1]{Nuovo Cimento {\bf #1}}
\newcommand{\CMP}[1]{Commun.\ Math.\ Phys.\ {\bf #1}}
\newcommand{\PR}[1]{Phys.\ Rev.\ {\bf #1}}
\newcommand{\PRL}[1]{Phys.\ Rev.\ Lett.\ {\bf #1}}
\newcommand{\MPL}[1]{Mod.\ Phys.\ Lett.\ {\bf #1}}
\newcommand{\BLMS}[1]{Bull.\ London Math.\ Soc.\ {\bf #1}}
\newcommand{\IJMP}[1]{Int.\ J.\ Mod.\ Phys.\ {\bf #1}}
\newcommand{\JMP}[1]{Jour.\ Math.\ Phys.\ {\bf #1}}
\newcommand{\LMP}[1]{Lett.\ Math.\ Phys.\ {\bf #1}}
\newpage
\setcounter{page}{0} \pagestyle{empty} \vs{12}
\begin{center}
{\LARGE {\bf Division Algebras and}}\\ {\quad}\\ {\LARGE{\bf
Extended $N=2,4,8$ SuperKdVs}}\\
 [0.8cm]

\vs{10} {\large H.L. Carrion, M. Rojas and F. Toppan} ~\\ \quad
\\
 {\large{\em CBPF - CCP}}\\{\em Rua Dr. Xavier Sigaud
150, cep 22290-180 Rio de Janeiro (RJ)}\\{\em Brazil}\\

\end{center}
{\quad}\\

\vs{6}

The first example of an $N=8$ supersymmetric extension of the KdV
equation is here explicitly constructed. It involves $8$ bosonic
and $8$ fermionic fields. It corresponds to the unique $N=8$
solution based on a generalized hamiltonian dynamics with
(generalized) Poisson brackets given by the Non-associative $N=8$
Superconformal Algebra. The complete list of inequivalent classes
of parametric-dependent $N=3$ and $N=4$ superKdVs obtained from
the ``Non-associative $N=8$ SCA" is also furnished. Furthermore, a
fundamental domain characterizing the class of inequivalent $N=4$
superKdVs based on the ``minimal $N=4$ SCA" is given.

\vs{6} \vfill \rightline{CBPF-NF-012/01} {\em E-mails:}{
hleny@cbpf.br; mrojas@cbpf.br; toppan@cbpf.br}
\pagestyle{plain}
\renewcommand{\thefootnote}{\arabic{footnote}}

\section{Introduction.}

In the last several years integrable hierarchies of non-linear
differential equations in $1+1$ dimensions have been intensely
explored, mainly in connection with the discretization of the
two-dimensional gravity (see \cite{dgz}).\par Supersymmetric
extensions of such equations have also been largely investigated
\cite{man}-\cite{gr} using a variety of different methods. Unlike the
bosonic theory, many questions have not yet been answered in the
supersymmetric case.\par In this paper we construct the first
example of a global $N=8$ supersymmetric extension of the KdV
equation. The strategy used is based on the derivation of the
supersymmetric non-linear equations from a generalized hamiltonian
system admitting the ``Non-Associative $N=8$ Superconformal
Algebra" of Englert et al. \cite{eng} as a generalized Poisson
bracket. The non-associativity of such an algebra  (i.e. the
failure in fulfilling the Jacobi identities) allows to overcome a
no-go theorem based on strict mathematical results. The
higher-derivative term in the KdV equation can be seen as induced
by the central extension of the Virasoro algebra. However, the
complete list of allowed central charges for (ordinary)
$N$-extended superconformal algebras has been produced in the
mathematical literature \cite{gls}. Central charges can be
introduced for $N\leq 4$ only. Indeed, supersymmetric
generalizations of KdV up to $N=4$ have been constructed
\cite{mat,di}. In order to construct supersymmetric generalizations
of KdV for $N>4$ one is therefore led to relax some condition on
the nature of the superconformal algebras of Poisson brackets.
Allowing non-associativity as in the $N=8$ SCA of reference
\cite{eng} makes possible to introduce a central extension. It is
therefore worth investigating whether this superconformal algebra
can be related to the construction of $N$-extended superKdVs
beyond the $N=4$ barrier. This is the purpose of the present
paper.
\par
The ``Non-Associative $N=8$ SCA" involves $8$ bosonic and $8$
fermionic fields and is constructed in terms of octonionic
structure constants. Its restriction to its real, complex or
quaternionic subalgebras leads, respectively, to the ordinary
$N=1,2,4$ Superconformal Algebras (in the last case it is the
so-called ``minimal $N=4$ SCA").\par In this paper at first we
revisit the $N=2,4$ KdV equations in the language of division
algebras. We construct a fundamental domain for the parametric
space of the inequivalent $N=4$ KdVs (our results complete and
complement the work of \cite{di}) and discuss the issue of
integrability.\par Later we apply the same techniques to
investigate the most general globally $N=8$ invariant generalized
hamiltonian for superextended KdV. It turns out that, if we
further assume invariance under octonionic involutions, the
hamiltonian is unique up to the normalization factor, giving rise
to a unique set of $N=8$ KdV equations. Such equations,
consistently reduced to the quaternionic subspace, produce the
most symmetric (global $SU(2)$-invariant) $N=4$ KdV set of
equations. This $N=4$ KdV system, despite being the most symmetric
one, does not correspond to the integrable point of $N=4$ KdV.
This result therefore suggests that the unique $N=8$ KdV is not an
integrable system.
\par On the other hand the authors of \cite{dgi} pointed out
that global $N=2$ supersymmetric systems can be obtained from the
``minimal $N=4$ SCA" Poisson brackets. We extended here such
analysis by investigating the class of global $N=3$ and $N=4$
supersymmetric extensions of KdV which can be constructed via the
``Non-Associative $N=8$ SCA" generalized Poisson brackets. The
complete solution is reported. In the $N=4$ case two inequivalent
classes (both parametric-dependent) of solutions, are found. The
existence of two $N=4$ classes is in consequence of the two
inequivalent ways of associating three invariant supersymmetry
charges with imaginary octonions (i.e. either producing, or not,
an $su(2)$ subalgebra), while the extra supersymmetry charge is
always associated with the octonionic identity. In the $N=3$ case
just a single class of parametric solutions is found since any
given pair of imaginary octonions is equivalent to any other
pair.\par We did not investigate here the issue of integrability
since our focus was in the construction of supersymmetric
extensions. However, we can notice that in the first class of
$N=4$ superKdV extension obtained from the ``Non-associative $N=8$
SCA" the parameters can be conveniently chosen so that a
consistent reduction to the integrable $N=4$ KdV can be made. This
leaves room to the possibility that the integrable $N=4$ KdV can
be embedded in such a larger $N=4$ system which still preserves
integrability.\par Some further comments are in order. This work
is partly a continuation of our previous one \cite{crt} concerning
the relation between the ``Non-Associative $N=8$ SCA" and the
superaffined octonionic algebra. Indeed, by reconstructing via
Sugawara the $N=8$ SCA fields with the affine fields, we can
induce on the affine fields a global $N=8$ set of equations,
generalizing both the NLS and mKdV equations, as well as the $N=4$
construction of reference \cite{ikt}.\par We heavily relied on the
Thielemans'package for computing classical OPE's with Mathematica
\cite{thi}, supported by our own package to deal with octonionic
structure constants.

\vspace{0.2cm}\noindent{\section{On Division Algebras and the
``Non-Associative $N=8$ SCA".}}

In this section we recall (see \cite{gk} and \cite{crt}) the
basic properties of the division algebra of the octonions which
will be used in the following and introduce the ``Non-Associative
$N=8$ Superconformal Algebra" according to \cite{eng} (see also
\cite{crt}).\par A generic octonion $x$ is expressed as
$x=x_a\tau_a$ (throughout the text the convention over repeated
indices, unless explicitly mentioned, is understood), where $x_a$
are real numbers while $\tau_a$ denote the basic octonions, with
$a=0,1,2,...,7$.\par $\tau_0\equiv {\bf 1}$ is the identity, while
$\tau_\alpha$, for $\alpha =1,2,...,7$, denote the imaginary
octonions. In the following a Greek index is employed for
imaginary octonions, a Latin index for the whole set of octonions
(identity included).\par The octonionic multiplication can be
introduced through
\begin{eqnarray}
\tau_\alpha \cdot \tau_\beta &=& -\delta_{\alpha\beta} \tau_0 +
C_{\alpha\beta\gamma} \tau_\gamma ,
\end{eqnarray}
with $C_{\alpha\beta\gamma}$ a set of totally antisymmetric
structure constants which, without loss of generality, can be
taken to be
\begin{eqnarray}
&C_{123}=C_{147}=C_{165}=C_{246}=C_{257}= C_{354}=C_{367}=1.&
\label{triple}
\end{eqnarray}
and vanishing otherwise.\par It is also convenient to introduce,
in the seven-dimensional imaginary octonions space, a $4$-indices
totally antisymmetric tensor $C_{\alpha\beta\gamma\delta}$, dual
to $C_{\alpha\beta\gamma}$, through \begin{eqnarray}
C_{\alpha\beta\gamma\delta} &=&
\frac{1}{6}\varepsilon_{\alpha\beta\gamma\delta\epsilon\zeta\eta}
C_{\epsilon\zeta\eta}
\end{eqnarray}
(the totally antisymmetric tensor
$\varepsilon_{\alpha\beta\gamma\delta\epsilon\zeta\eta}$ is
normalized so that $\varepsilon_{1234567}=+1$). \par The
octonionic multiplication is not associative since for generic
$a,b,c$ we get$(\tau_a\cdot \tau_b)\cdot \tau_c \neq \tau_a \cdot
(\tau_b\cdot\tau_c)$. However, the weaker condition of
alternativity is satisfied. This means that, for $a=b$, the
associator
\begin{eqnarray}
\relax [ \tau_a,\tau_b,\tau_c]\equiv (\tau_a\cdot\tau_b)\cdot
\tau_c -\tau_a\cdot(\tau_b\cdot \tau_c)
\end{eqnarray}
is vanishing.\par The specialization of the octonionic indices to,
let's say, $0,1$ or $0,1,2,3$ leads respectively to the complex
number or to the division algebra of quaternions.\par The
octonionic algebra admits seven involutions, specified by the
mappings
\begin{eqnarray}
& \tau_0\mapsto \tau_0,\quad\tau_p\mapsto \tau_p,\quad
\tau_q\mapsto -\tau_q,&
\end{eqnarray}
where $p$ takes value in one of the seven triples entering
(\ref{triple}), while $q$ specifies the four complementary values.
The three involutions for the quaternions (with two generators)
are recovered as the restrictions to the $0,1,2,3$ subspace.\par
The $N=8$ extension of the Virasoro algebra (Non-associative $N=8$
SCA) involves $8$ bosonic and $8$ fermionic fields and is
constructed in terms of the octonionic structure constants.
Besides the spin-$2$ Virasoro field denoted as $T$, it contains
eight fermionic spin-$\frac{3}{2}$ fields $Q$, $Q_\alpha$ and $7$
spin-$1$ bosonic currents $J_\alpha$. It is explicitly given by
the following Poisson brackets
\begin{eqnarray}
\{ T(x), T(y)\} &=& -\frac{1}{2} {\partial_y}^3 \delta(x-y) +
2T(y)\partial_y\delta(x-y) +T'(y) \delta(x-y),\nonumber\\ \{T(x),
Q(y)\} &=& \frac{3}{2}Q(y)\partial_y\delta (x-y) + Q'(y) \delta
(x-y),\nonumber\\ \{T(x), Q_\alpha(y)\} &=&
\frac{3}{2}Q_{\alpha}(y)\partial_y\delta (x-y) + {Q_\alpha}'(y)
\delta (x-y),\nonumber\\
 \{T(x), J_\alpha(y)\} &=&
J_{\alpha}(y)\partial_y\delta (x-y) + {J_\alpha}'(y) \delta
(x-y),\nonumber\\
 \{Q(x), Q(y)\} &=&-\frac{1}{2}{\partial_y}^2\delta(x-y)+
+\frac{1}{2} {T}(y) \delta (x-y),\nonumber\\
 \{Q(x), Q_\alpha(y)\} &=&
-J_{\alpha}(y)\partial_y\delta (x-y) -\frac{1}{2} {J_\alpha}'(y)
\delta (x-y),\nonumber\\
 \{Q(x), J_\alpha(y)\} &=&
-\frac{1}{2}Q_{\alpha}(y)\delta (x-y),\nonumber\\
 \{Q_\alpha(x), Q_\beta(y)\}
 &=&-\frac{1}{2}\delta_{\alpha\beta}{\partial_y}^2\delta(x-y) +
 C_{\alpha\beta\gamma}
 J_\gamma(y)\partial_y\delta(x-y)+\nonumber\\&& +
\frac{1}{2}(\delta_{\alpha\beta}T(y)+C_{\alpha\beta\gamma}
{J_\gamma}'(y))\delta(x-y),\nonumber\\
 \{Q_\alpha(x), J_\beta(y)\} &=&
\frac{1}{2}(\delta_{\alpha\beta} Q(y)-C_{\alpha\beta\gamma}
Q_\gamma (y)) \delta(x-y),\nonumber\\
 \{J_\alpha(x), J_\beta(y)\} &=&
\frac{1}{2}\delta_{\alpha\beta}\partial_y\delta (x-y) -
C_{\alpha\beta\gamma} J_\gamma(y)\delta (x-y). \label{N8SCA}
\end{eqnarray}
Notice the presence of the central term, essential in order to
obtain supersymmetric KdV equations. Due to the non-associativity
of octonions the structure constants of (\ref{N8SCA}) do not satisfy
the Jacobi identity (see \cite{crt} for a detailed discussion).

\vspace{0.2cm}\noindent{\section{The $N=2$ and the $N=4$ KdVs
Revisited.}}

By restricting the Greek indices to take either the values $1$ or
$1,2,3$, we recover from (\ref{N8SCA}) the $N=2$ and the $N=4$
Superconformal algebras respectively (in the case of $N=4$ the
corresponding algebra is known as the ``minimal $N=4$ SCA"). They
can be regarded as one of the Poisson brackets for the $N=2$ and
the $N=4$ KdVs \cite{mat,di}.\par These non-linear equations can be constructed
by looking for the most general hamiltonian with the right
dimension (i.e. whose hamiltonian density has dimension $4$)
invariant under global supersymmetric charges given by $\int dx
Q(x)$ and $\int dx Q_\alpha(x)$. This approach was used to
construct the $N=2$ KdV in \cite{mat}, while the $N=4$ KdV was
obtained in terms of a harmonic superspace formalism in
\cite{di}.\par For what concerns the $N=2$ case we summarize here
the results of \cite{mat}. We avoid writing explicit formulas
since they can be immediately recovered from a suitable reduction
of the $N=4$ KdV results as discussed later. Up to a normalization
factor, the $N=2$-invariant hamiltonians depend on a single real
parameter, denoted as ``$a$", which labels inequivalent $N=2$
KdVs. Three special values for $a$, i.e. $a=-2,1,4$, correspond to
the three inequivalent $N=2$ KdV equations which are integrable.
The integrability for these special values of $a$ was at first
suggested (and proven for $a=-2,4$)  in \cite{mat} after checking the existence of higher
order hamiltonians in involution among themselves and with respect
to  the original $a$-dependent $N=2$-invariant one. Later the
integrability of $a=1$ was proven in the first reference of \cite{pop} with the explicit
construction of the corresponding Lax operators.\par Here we
extend the analysis of \cite{mat} to the $N=4$ KdV case. In
particular we are able to fully determine the moduli space of
inequivalent $N=4$ KdVs. Our results extend and complete those
originally appeared in \cite{di}.\par The most general
$N=4$-invariant hamiltonian of right dimension depends on $5$
parameters (apart the overall normalization factor) and is
explicitly given by
\begin{eqnarray}
\relax H&=&\int dx
[-2 T^2 -2Q'Q-2Q_\alpha'Q_\alpha +2J_\alpha''J_\alpha +x_\alpha T{J_\alpha}^2 +
 2x_\alpha
QQ_\alpha J_\alpha -\epsilon_{\alpha\beta\gamma}x_\gamma
Q_\alpha Q_\beta J_\gamma+\nonumber\\&&
\frac{1}{3}\epsilon_{\alpha\beta\gamma} (x_\beta-x_\alpha)
J_\alpha J_\beta{J_\gamma}'
-2z_\alpha\epsilon_{\alpha\beta\gamma}TJ_\beta J_\gamma
-
\nonumber\\
&& 2z_1 Q(Q_2J_3+Q_3J_2)-2z_2Q(Q_3J_1+Q_1J_3)-2z_3Q(Q_1J_2+Q_2J_1)+\nonumber\\
&&
2 z_1Q_1 (Q_2J_2-Q_3J_3)
+2 z_3Q_3 (Q_1J_1-Q_2J_2)
+2 z_2Q_2 (Q_3J_3-Q_1J_1)
-\nonumber\\&&
z_1{J_1}'({J_2}^2-{J_3}^2)
-z_3{J_3}'({J_1}^2-{J_2}^2)
-z_2{J_2}'({J_3}^2-{J_1}^2)
],
\end{eqnarray}
where the convention over repeated indices is understood and $\alpha,\beta,\gamma$
are restricted
to $1,2,3$, while $\epsilon_{123}=1$.\par
In order to guarantee the $N=4$ invariance the three parameters $x_\alpha$
must satisfy the condition
\begin{eqnarray}
x_1+x_2+x_3&=&0,
\label{condition}
\end{eqnarray}
so that only two of them are truly independent (together with the three $z_\alpha$'s
they provide the five parameters mentioned above).
However, the further requirement for the
hamiltonian to be invariant not only under global $N=4$
supersymmetry, but also under the three involutions of the $N=4$
Superconformal Algebra (obtained by flipping the sign of the four
fields $J_\alpha$, $Q_\alpha$, for $\alpha =1,2$, $\alpha=1,3$ and
$\alpha=2,3$ respectively, while leaving unchanged the remaining
four fields) kills the three $z_\alpha$'s parameters, which must be set equal to zero.
\par The most
general hamiltonian of such a kind is therefore given by
\begin{eqnarray}
\relax H&=&\int dx
[-2T^2 -2Q'Q-2Q_\alpha'Q_\alpha +2J_\alpha''J_\alpha +x_\alpha T{J_\alpha}^2 +
 2x_\alpha
QQ_\alpha J_\alpha -\epsilon_{\alpha\beta\gamma}x_\gamma
Q_\alpha Q_\beta J_\gamma +\nonumber\\&&
\frac{1}{3}\epsilon_{\alpha\beta\gamma} (x_\beta-x_\alpha)
J_\alpha J_\beta{J_\gamma}' ].\label{ham}
\end{eqnarray}
where of course (\ref{condition}) continues to hold.\par Since any given
ordered pair of the three parameters $x_\alpha$ can be chosen to
be plotted along the $x$ and $y$ axis describing a real $x-y$
plane, it can be easily proven that the fundamental domain of the
moduli space of inequivalent $N=4$ KdV equations can be chosen to
be the region of the plane comprised between the real axis $y=0$
and the $y=x$ line (boundaries included). Five other regions of
the plane (all such regions are related via an $S_3$-group
transformation) could as well be chosen as the fundamental
domain.\par In the region of our choice, the $y=x$ line
corresponds to an extra global $U(1)$-invariance, since the
hamiltonian whose parameters live in this line is in involution
with the global charge $\int dx J_3$ (namely $\{ H, \int dx\cdot
J_3\}=0$). The origin, that is $x_1=x_2=x_3=0$, is the most
symmetric point, corresponding to a global $SU(2)$ invariance, the
given hamiltonian being in involution with respect to the three
$\int dx\cdot J_\alpha$ charges.
\par
The equations of motion for the whole class of inequivalent $N=4$ KdV's are
given by
\begin{eqnarray}
{\dot T}&=& -T'''-12T'T -6Q''Q-6{Q_{\alpha}}''Q_\alpha
+(4+\frac{x_\alpha}{2})J_\alpha'''J_\alpha +\frac{3}{2}x_\alpha
{J_\alpha}''{J_\alpha}' + 3x_\alpha (T{J_\alpha}^2)' +
\nonumber\\&& 6x_\alpha (QQ_\alpha J_\alpha)'
-3x_\gamma\epsilon_{\alpha\beta\gamma} (Q_\alpha Q_\beta
J_\gamma)'+\epsilon_{\alpha\beta\gamma}
(x_\gamma-x_\beta)({J_\alpha}''J_\beta J_\gamma
-J_\alpha{J_\beta}'J_{\gamma}'),\nonumber\\  {\dot Q}&=&
-Q'''-6(TQ)'-(4+\frac{x_\alpha}{2})({Q_\alpha}'J_\alpha)'+
(2-\frac{x_\alpha}{2}) (Q_\alpha {J_\alpha}')'+3x_\alpha
(Q{J_\alpha}^2)'-\nonumber\\&&
\epsilon_{\alpha\beta\gamma}(x_\gamma-x_\beta)(Q_\alpha J_\beta
J_\gamma)',\nonumber\\ {\dot Q_\alpha} &=& -{Q_{\alpha}}'''
-6(TQ_\alpha )'+(4+\frac{x_\alpha}{2}) (Q'J_\alpha)'
-(2-\frac{x_\alpha}{2}) (Q{J_\alpha}')'+3x_\beta (Q_\alpha
{J_\beta}^2)'+\nonumber\\ && \epsilon_{\alpha\beta\gamma}
(x_\gamma-x_\beta)(QJ_\beta
J_\gamma)'+\epsilon_{\alpha\beta\gamma}(4+\frac{x_\gamma}{2})
({Q_\beta}'
J_\gamma)'-\epsilon_{\alpha\beta\gamma}(2-\frac{x_\gamma}{2})(Q_\beta
{J_\gamma}')'+\nonumber\\ && 2(x_\beta-x_\alpha)
(1-\delta_{\alpha\beta})(J_\alpha Q_\beta J_\beta)',\nonumber\\
{\dot{J_\alpha}} &=&
-{J_\alpha}'''-(4+\frac{x_\alpha}{2})(TJ_\alpha)'+(2-\frac{x_\alpha}{2})(QQ_\alpha)'-
2(x_\alpha+x_\beta)Q_\alpha Q_\beta  J_\beta-\nonumber\\ &&
\epsilon_{\alpha\beta\gamma}(1-\frac{x_\alpha}{4}) (Q_\beta
Q_\gamma)'-2\epsilon_{\alpha\beta\gamma} (x_\gamma - x_\beta) Q
Q_\beta J_\gamma +\epsilon_{\alpha \beta\gamma}
(4+\frac{x_\gamma}{2})({J_\beta}'J_\gamma)'+\nonumber\\ && 3
x_\beta {J_\alpha}'{J_\beta}^2
+2(1-\delta_{\alpha\beta})(x_\beta-x_\alpha)J_\alpha
{J_\beta}'J_\beta .
\end{eqnarray} where the constraint $x_1+x_2+x_3=0$ is satisfied
and $(x_1,x_2)$ take value either in the region $I \equiv \{x_1,
x_2| x_2\geq x_1\geq 0\}$ or in $II\equiv \{x_1,x_2|x_2\leq
x_1\leq 0\}$. Each given pair $(x_1,x_2) \in I\cup II$ labels an
inequivalent $N=4$ KdV equation.\par The three involutions (each
one associated to any given imaginary quaternion) allows to
perform three consistent reduction of the $N=4$ KdV equation to an
$N=2$ KdV, by setting simultaneously equal to $0$ all the fields
associated with the $\tau$'s which flip the sign (confront the
discussion in the previous section). Therefore the first
involution allows to consistently set equal to zero the fields
$J_2=J_3=Q_2=Q_3=0$, leaving the $N=2$ KdV equation for the
surviving fields $T, Q, Q_1, J_1$. Similarly, the second and the
third involution allows to set equal to zero the four fields
labeled by $1,3$ and $1,2$ respectively. It turns out that to each
such reduction only one free parameter survives, namely $x_1$,
$x_2$ or respectively $x_3$.\par This remaining free parameter
coincides up to a normalization factor with the free parameter $a$
of reference \cite{mat}. More specifically
\begin{eqnarray}
a&=& \frac{1}{4}x_\alpha
\end{eqnarray}
with $\alpha=1,2,3$ according to the reduction.\par As a
consequence, a necessary condition for the integrability of the
$N=4$ KdV requires that for a given pair $(x_1,x_2)\in I\cup II$
each one of the three reductions produce for $a$ one of the known
integrable values of $a$, namely $-2,1,4$. It is then easily
checked that there are only two points in $I\cup II$, both in the
$U(1)$-invariant $x_1=x_2$ line, implying integrability for the
three reduced $N=2$ KdV's. The solutions are\\ {\em i}) $x_1=x_2=
-8$, ($x_3 = 16$) and \\ {\em ii}) $x_1=x_2=4$, ($x_3= -8$).
\par
The first point, which produces the $a=-2$ and the $a=4$
integrable $N=2$ KdV's after reduction, is the integrable point
discussed in \cite{di}. For what concerns
the second point, despite the fact that it allows the reduction to
the $a=1$ and the $a=-2$ integrable $N=2$ KdV's, it does not seem
to correspond to an $N=4$ integrable hierarchy. We explicitly
constructed the most general global $N=4$ and $U(1)$ invariant
hamiltonian whose hamiltonian density has total dimension
dimension $6$. This would correspond to the third hamiltonian in
the KdV hierarchy ((\ref{ham}) would be the second hamiltonian).
This hamiltonian however fails to be compatible with the third
hamiltonian of the corresponding integrable $N=2$ KdV's. More
precisely, the three (two independent) reductions to $N=2$
produce hamiltonians which should coincide with the third
hamiltonian of the $N=2$ KdV for the corresponding value of $a$.
While this is true for the first solution ($x_1=x_2=-8$,
$x_3=16$), this is no longer true for the second choice of values
($x_1=x_2=4$, $x_3=-8$), as we explicitly verified.\par This
computation does not yet rule out the possibility that {\em ii})
would be a point of integrability for the $N=4$ KdV. It would
still be possible that it corresponds to an integrable hierarchy
with a ``missed" hamiltonian for the hamiltonian density of dimension $6$.
\par
The origin ($x_1=x_2=x_3=0$) corresponds to the most symmetric point,
being associated
to a global $SU(2)$ invariance, as already remarked. In any case it does
not correspond to an
integrable point of the $N=4$ KdV since its reductions to $N=2$ KdV
do not lead to one of the three integrable values of $a$.

\section{The $N=8$ SuperKdV.}

In this section we construct the first example of an $N=8$
supersymmetric extension of the KdV equation. In order to be able
to realize an $N=8$ KdV we extend the method discussed in the
previous section to the case of the ``Non-Associative $N=8$
Superconformal Algebra" (\ref{N8SCA}). The reason why we are forced
to make use of a non-associative algebra has been discussed in the
Introduction.\par More specifically, we started with the most
general hamiltonian of right dimension (its hamiltonian density
having dimension equal to $4$) constructed with the $16$ ($8$
bosonic and $8$ fermionic) fields entering (\ref{N8SCA}). Later we
imposed some constraints on it. At first we restricted the free
coefficients in order to make the resulting hamiltonian invariant
under the whole set of seven involutions of the $N=8$
superconformal algebra. This is the $N=8$ extension of
a requirement already encountered in the $N=4$ case. The seven
involutions are so defined. The fields $T, Q$ are unchanged, as
well as the $6$ fields $Q_\alpha,J_\alpha$, for the $\alpha$'s
taking value in one of the seven triples entering (\ref{triple}). The
$8$ remaining fields $Q_\beta, J_\beta$, with $\beta$ labeling the
four complementary values (for any given choice of the original
triple), have the sign flipped ($Q_\beta\mapsto -Q_\beta$,
$J_\beta\mapsto - J_\beta$). After having constructed the most
general hamiltonian $H$ invariant under the whole set of seven
involutions, we started imposing the invariance under the $N=8$
global supersymmetric transformations, that is we required
\begin{eqnarray}
\{ \int dx\cdot Q_a(x), H\} &=& 0,
\end{eqnarray}
for $a=0,1,2,...,7$ (here $Q_0\equiv Q$), while $\{\star,\star\}$
denotes the generalized Poisson brackets given by the
Non-associative $N=8$ SCA (\ref{N8SCA}).\par It is worth to point
out that for this generalized hamiltonian system, the Poisson
brackets are assumed to be classical. In particular they satisfy
the Leibniz property (or, better, its graded version due to the
supersymmetry of (\ref{N8SCA})). The only feature of the
non-associativity of the octonions lie in the non-vanishing of the
Jacobi identities for the structure constants of the (\ref{N8SCA})
algebra. The fields entering (\ref{N8SCA}) are assumed to be
ordinary (bosonic and fermionic) real fields.\par  Needless to
say, the get the final answer we heavily relied on Mathematica's
computations for classical OPE's, based both on the Thielemans'
package \cite{thi} and on our own package to deal with octonionic structure
constants.
\par
The final result is the following. There exists a unique hamiltonian
which is invariant
under the whole set of global $N=8$ supersymmetries. It admits no free
parameter (apart the trivial normalization factor) and is quadratic on the
fields. It is explicitly given by
\begin{eqnarray}
\relax H &=& \int dx [-2T^2 -2Q'Q-2Q_\alpha'Q_\alpha +2 J_\alpha''J_\alpha],
\label{hamn8}
\end{eqnarray}
(here $\alpha=1,2,...,7$ and the summation over repeated indices is understood).
This result implies that there is only one $N=4$ KdV system which can be
consistently extended to $N=8$ KdV, namely the one which corresponds to the
origin of the coordinates ($x_1=x_2=x_3=0$), that is the most symmetric point.
While the corresponding hamiltonian for the $N=4$ case admits a global $SU(2)$-invariance,
the $N=8$
hamiltonian (\ref{hamn8}) is invariant with respect to each one of the seven global charges
$\int dx \cdot J_\alpha (x)$, that is
\begin{eqnarray}
\{ \int dx \cdot J_\alpha (x), H\}&=& 0.
\end{eqnarray}
The seven charges $\int dx \cdot J_\alpha (x)$ generates a
symmetry which extends $SU(2)$; it does not correspond to a group
due to the non-associative character of the octonions.\par Despite
the apparent simplicity and the fact that it is quadratic in the
fields, the hamiltonian (\ref{hamn8}) generates an $N=8$
supersymmetric extension of KdV which is not integrable.
Better stated, even its $N=4$ KdV reduction does not correspond to an
integrable point of the $N=4$ KdV.\par
The equations of motion of the $N=8$ KdV are obtained through
\begin{eqnarray}
\dot{\Phi_i}&=& \{ \Phi_i, H\},\label{eom}
\end{eqnarray}
where $\Phi_i$ collectively denote the fields entering (\ref{N8SCA}).
\par
We explicitly obtain
\begin{eqnarray}
{\dot T} &=&- T''' - 12 T' T - 6 Q_a'' Q_a +4
J_\alpha''J_\alpha,\nonumber\\ {\dot Q} &=& -Q''' -6 T' Q -6 T Q' -
4 Q_\alpha''J_\alpha +2 Q_\alpha J_\alpha''-2Q_\alpha'J_\alpha',\nonumber\\ {\dot
Q}_\alpha &=& -{Q_\alpha}''' - 2 Q J_\alpha'' - 6 TQ_\alpha' - 6 T'
Q_\alpha + 2 Q' J_\alpha' +4 Q''J_\alpha - \nonumber\\ &&
2 C_{\alpha\beta\gamma} ( Q_\beta J_\gamma'' -Q_\beta' J_\gamma' -2
Q_\beta''J_\gamma ),\nonumber\\ {\dot J}_\alpha &=& -{J_\alpha}'''
- 4 T'J_\alpha - 4 T J_\alpha ' + 2 Q Q_\alpha '+2Q'Q_\alpha
-C_{\alpha\beta\gamma}( 4 J_\beta J_\gamma'' +2Q_\beta
Q_\gamma'). \label{EOMN8}
\end{eqnarray}
It is a simple exercise to prove that the equations of motion
(\ref{EOMN8}) are compatible with the $N=8$ global supersymmetries
generated by $\int dx \cdot Q_a(x)$ ($a=0,1,2,...,7$) which provide
the transformations
\begin{eqnarray}
\delta_a \Phi_i (y) &=& \{ \int dx \cdot Q_a(x), \Phi_i(y)\}.
\end{eqnarray}
The above system of equations corresponds to the first known
example of an $N=8$ supersymmetric extension of KdV.

\section{On Global $N=3$ and $N=4$ Extended SuperKdVs Based On the $N=8$ SCA.}

The authors of \cite{dgi} proved the existence of integrable
systems, obtained in terms of the $N=4$ Superconformal algebra,
which admit only an $N=2$ global supersymmetry.\par It is worth
considering in our context, which involves a larger number of
supersymmetries, which kind of extended supersymmetric systems are
supported by the Non-associative $N=8$ SCA. We present the
complete analysis of the $N=3$ and the $N=4$ solutions. We
construct the most general $N=3$ and $N=4$ superextensions of KdV
admitting the Non-associative $N=8$ SCA as generalized Poisson
brackets. Both such cases turn out to be parametric-dependent.\par
Apart the unique $N=8$ solution, $N=4$ is the largest number of
supersymmetries which can be consistently imposed (by assuming an
$N>4$ invariance we automatically recover the full $N=8$
invariance).\par Both in the $N=3$ and the $N=4$ cases, without
loss of generality, one of the invariant supersymmetric charges
can always be assumed to be $\int dx Q(x)$, with $Q(x)$ entering
(\ref{N8SCA}). In the $N=3$ case the two remaining invariant
supersymmetric charges (associated with imaginary octonions) can
be chosen at will, since all pairs of imaginary octonions are
equivalent. In the formula below, without loss of generality, we
chose the invariant supersymmetric charges being given by $\int dx
Q_1(x)$ and $\int dx Q_2(x)$.\par The situation is different in
the $N=4$ case. Now we have three extra invariant supersymmetric
charges to be associated with imaginary octonions. However, two
inequivalent ways in choosing a triple of imaginary octonions exist, depending
on whether the chosen triple corresponds to one of the seven values in (\ref{triple})
(i.e. the triples associated to an $su(2)$ subalgebra), or not. Two
inequivalent classes of solutions, labelled by $N=4$ $(I)$ and $N=4$ $(II)$ are
respectively obtained. The first $(I)$ class can be individuated by
choosing, without loss of generality, the three extra supersymmetric charges
to be given by $\int dx Q_1(x)$, $\int dx Q_2(x)$ and $\int dx Q_3(x)$. The second
class $(II)$, without loss of generality, can be produced by assuming invariance
under $\int dx Q_1(x)$, $\int dx Q_2(x)$ and $\int dx Q_4(x)$.\par
Let us present now the complete solutions.\par
The most general $N=3$ invariant hamiltonian depends (up to the normalization
factor) on $6$ free parameters entering $x$ and $x_\tau$ ($\tau=1,2,...,7$).\par
The seven $x_\tau$'s satisfy two constraints
\begin{eqnarray}
x_1+x_2+x_3&=&0,\nonumber\\
x_4+x_5+x_6+x_7&=& 0.
\end{eqnarray}
The most general hamiltonian is given by
\begin{eqnarray}
 \relax
H&=&\int dx [-2T^2 -2Q'Q-2 Q_\alpha'Q_\alpha +x{Q_\mu}'Q_\mu
+2J_\alpha''J_\alpha-x{J_\mu}''J_\mu +x_\alpha T{J_\alpha}^2 +
x_\mu T {J_\mu}^2+\nonumber\\&&
 2x_\alpha
QQ_\alpha J_\alpha+2x_\mu Q Q_\mu J_\mu-  x_\gamma
C_{\alpha\beta\gamma} Q_\alpha Q_\beta J_\gamma -x_\nu
C_{\alpha\mu\nu} Q_\alpha Q_\mu J_\nu +\nonumber\\&&(x_\mu+x_\nu)
C_{\mu\nu\alpha} Q_\mu Q_\nu J_\alpha +\frac{1}{3}
C_{\alpha\beta\gamma} (x_\beta-x_\alpha) J_\alpha J_\beta
{J_\gamma}' + 2x_\mu C_{\alpha\mu\nu} J_\alpha J_\mu {J_\nu}'
 ].\label{hamn3}
\end{eqnarray}
where $\alpha,\beta,\gamma$ are restricted to take the values $1,2,3$, while
$\mu,\nu$ are restricted to the complementary values $4,5,6,7$.\par
The equations of motion for this $N=3$ generalization of KdV are directly
computed from (\ref{hamn3}) by applying the Poisson brackets, like in (\ref{eom}).\par
The complete set of equations is written down in $37$ pages of LaTex. For that reason
they are not being reported here. The corresponding LaTex file however is available upon
request.\par
For what concerns the $N=4$ cases, the $(I)$ class of solutions involve three free
parameters (up to the normalization factor) entering $x$ and $x_\alpha$ ($\alpha=1,2,3$),
where the $x_\alpha$'s are constrained to satisfy $x_1+x_2+x_3=0$.
\par
The most general $N=4$ -invariant hamiltonian of type $(I)$ is given by
\begin{eqnarray}
 \relax
H&=&\int dx [-2T^2 -2Q'Q-2 Q_\alpha'Q_\alpha +x{Q_\mu}'Q_\mu
+2J_\alpha''J_\alpha-x{J_\mu}''J_\mu +x_\alpha T{J_\alpha}^2 +
x_\mu T {J_\mu}^2+\nonumber\\&&
 2x_\alpha
QQ_\alpha J_\alpha+2x_\mu Q Q_\mu J_\mu-  x_\gamma
C_{\alpha\beta\gamma} Q_\alpha Q_\beta J_\gamma -x_\nu
C_{\alpha\mu\nu} Q_\alpha Q_\mu J_\nu +\nonumber\\&&(x_\mu+x_\nu)
C_{\mu\nu\alpha} Q_\mu Q_\nu J_\alpha +\frac{1}{3}
C_{\alpha\beta\gamma} (x_\beta-x_\alpha) J_\alpha J_\beta
{J_\gamma}' + 2x_\mu C_{\alpha\mu\nu} J_\alpha J_\mu {J_\nu}'
 ].\label{(I)}
\end{eqnarray}
As before $\alpha,\beta,\gamma=1,2,3$, while $\mu,\nu$ take the values  $4,5,6,7$.\par
The $N=4$ $(I)$ equations of motion are explicitly given by
\begin{eqnarray}
{\dot T}&=& -T'''-12T'T -6Q''Q-6{Q_{\alpha}}''Q_\alpha
+(4+\frac{x_\alpha}{2})J_\alpha'''J_\alpha +\frac{3}{2}x_\alpha
{J_\alpha}''{J_\alpha}'+3x{Q_\mu}''Q_\mu -2x {J_\mu}'''J_\mu
+\nonumber\\&& 3x_\alpha (T{J_\alpha}^2)' +  6x_\alpha (QQ_\alpha
J_\alpha)' -3x_\gamma C_{\alpha\beta\gamma} (Q_\alpha Q_\beta
J_\gamma)'+C_{\alpha\beta\gamma}
(x_\gamma-x_\beta)({J_\alpha}''J_\beta J_\gamma
-J_\alpha{J_\beta}'J_{\gamma}'),\nonumber\\ {\dot Q}&=&
-Q'''-6(TQ)'-(4+\frac{x_\alpha}{2})({Q_\alpha}'J_\alpha)'+
(2-\frac{x_\alpha}{2}) (Q_\alpha {J_\alpha}')'+ 2x
({Q_\mu}'J_\mu)'-x(Q_\mu{J_\mu}')' +\nonumber\\&& 3x_\alpha
(Q{J_\alpha}^2)'- C_{\alpha\beta\gamma}(x_\gamma-x_\beta)(Q_\alpha
J_\beta J_\gamma)',\nonumber\\ {\dot Q_\alpha} &=&
-{Q_{\alpha}}''' -6(TQ_\alpha )'+(4+\frac{x_\alpha}{2})
(Q'J_\alpha)' -(2-\frac{x_\alpha}{2}) (Q{J_\alpha}')'+3x_\beta
(Q_\alpha {J_\beta}^2)'-\nonumber\\ && 2x
C_{\alpha\mu\nu}({Q_\mu}'J_\nu)'+x
C_{\alpha\mu\nu}(Q_\mu{J_\nu}')' +C_{\alpha\beta\gamma}
(x_\gamma-x_\beta)(QJ_\beta J_\gamma)'+\nonumber\\&&
C_{\alpha\beta\gamma}(4+\frac{x_\gamma}{2}) ({Q_\beta}'
J_\gamma)'-C_{\alpha\beta\gamma}(2-\frac{x_\gamma}{2})(Q_\beta
{J_\gamma}')'+ 2(x_\beta-x_\alpha)
(1-\delta_{\alpha\beta})(J_\alpha Q_\beta J_\beta)',\nonumber\\
{\dot {Q_\mu}}&=&
\frac{x}{2}{Q_\mu}'''+(x-4)T{Q_\mu}'-6T'Q_\mu+4Q''J_\mu+2Q'{J_\mu}'+xQ{J_\mu}''+
4C_{\mu\alpha\nu}{Q_\alpha}''J_\nu+2C_{\mu\alpha\nu}{Q_\alpha}'{J_\nu}'+\nonumber\\
&& xC_{\mu\alpha\nu}Q_\alpha{J_\nu}''
-2xC_{\mu\nu\alpha}{Q_\nu}''J_\alpha-xC_{\mu\nu\alpha}
{Q_\nu}'{J_\alpha}' -2C_{\mu\nu\alpha}Q_\nu{J_\alpha}'' +x_\alpha
C_{\mu\alpha\nu}Q Q_\alpha Q_\nu-\nonumber\\&& x_\alpha
C_{\mu\alpha\nu}Q J_\alpha {J_\nu}'-2x_\alpha C_{\mu\alpha\nu}Q
{J_\alpha}'J_\nu-2x_\alpha C_{\mu\alpha\nu} Q'J_\alpha J_\nu
+2x_\alpha {Q_\alpha}'J_\alpha J_\mu+x_\alpha
{Q_\mu}'{J_\alpha}^2+\nonumber\\&& 3x_\alpha Q_\mu
{J_\alpha}'J_\alpha-x_\alpha C_{\mu\nu\alpha} TQ_\nu J_\alpha +
x_\alpha Q_\alpha J_\alpha {J_\mu}' +2x_\alpha Q_\alpha {J_\alpha
}'J_\mu + \frac{1}{2} C_{\mu\alpha\beta\nu}
(x_\alpha+x_\beta)Q_\alpha Q_\beta Q_\nu -\nonumber\\&&2x_\beta
C_{\mu\alpha\beta\nu} {Q_\alpha}'J_\beta J_\nu +x_\alpha
C_{\mu\nu\alpha\beta} Q_\nu J_\alpha {J_\beta}'-2x_\beta
C_{\mu\alpha\beta\nu} Q_\alpha {J_\beta}'J_\nu -x_\beta
C_{\mu\alpha\beta\nu}Q_{\alpha} J_\beta{J_\nu}' ,
 \nonumber\\{\dot{J_\alpha}}
&=&
-{J_\alpha}'''-(4+\frac{x_\alpha}{2})(TJ_\alpha)'+(2-\frac{x_\alpha}{2})(QQ_\alpha)'-
2(x_\alpha+x_\beta)Q_\alpha Q_\beta  J_\beta-\nonumber\\ &&
C_{\alpha\beta\gamma}(1-\frac{x_\alpha}{4}) (Q_\beta Q_\gamma)'+x
C_{\alpha\mu\nu}{Q_\mu}'Q_\nu -2C_{\alpha\beta\gamma} (x_\gamma -
x_\beta) Q Q_\beta J_\gamma +C_{\alpha \beta\gamma}
(4+\frac{x_\gamma}{2})({J_\beta}'J_\gamma)'-\nonumber\\&&
2 x C_{\alpha\mu\nu}{J_\mu}''J_\nu+ 3 x_\beta {J_\alpha}'{J_\beta}^2
+2(1-\delta_{\alpha\beta})(x_\beta-x_\alpha)J_\alpha
{J_\beta}'J_\beta,\nonumber\\ {\dot{J_\mu}} &=& \frac{1}{2}x
{J_\mu}'''-4(TJ_\mu)'+2Q'Q_\mu-xQ{Q_\mu}'
-2C_{\mu\alpha\nu}{Q_\alpha}'Q_\nu +xC_{\mu\alpha\nu} Q_\alpha
{Q_\nu}'-4C_{\mu\nu\alpha} {J_\nu}''J_\alpha+\nonumber\\&& 2x
C_{\mu\alpha\nu}J_\alpha{J_\nu}''+2x_\alpha C_{\mu\alpha\nu} T
J_\alpha J_\nu -x_\alpha C_{\mu\nu\alpha}Q Q_\nu
J_\alpha+2x_\alpha C_{\mu\alpha\nu}QQ_\alpha J_\nu +x_\alpha
Q_\alpha J_\alpha Q_\mu +\nonumber\\&& x_\alpha
{J_\mu}'{J_\alpha}^2+ 2x{J_\alpha}'J_\alpha J_\mu + 2x_\alpha
C_{\mu\alpha\beta\nu}J_{\alpha}{J_\beta}'J_\nu +x_\beta
C_{\mu\alpha\beta\nu}Q_\alpha J_\beta Q_\nu +(x_\alpha+x_\beta)
C_{\mu\alpha\beta\nu} Q_\alpha Q_\beta J_\nu .\nonumber\\&&
\end{eqnarray}
The second $(II)$ class of $N=4$ solutions is two-parametric. The free
parameters can be chosen to be $x_1$ and $x_2$, while the remaining
$x_\tau$ parameters entering the hamiltonian below
are restricted to be
\begin{eqnarray}
x_3=x_4&=& -(x_1+x_2),\nonumber\\
x_5&=&0,\nonumber\\
x_6&=& x_1,\nonumber\\
x_7&=& x_2.
\end{eqnarray}
The most general $N=4$ $(II)$ hamiltonian is given by
\begin{eqnarray} \relax
H&=&\int dx [-2T^2 -2Q'Q-2 Q_\alpha'Q_\alpha +2J_\alpha''J_\alpha
+x_\alpha T{J_\alpha}^2 +
 2x_\alpha
QQ_\alpha J_\alpha+
 C_{\rho\sigma\lambda}(x_\rho +x_\sigma )
Q_\rho Q_\sigma J_\lambda
+\nonumber\\&&C_{\rho\lambda\sigma}(x_\rho+x_\lambda)Q_\rho
Q_\lambda J_\sigma -C_{\lambda \mu\nu} (x_\lambda + x_\mu)
Q_\lambda Q_\mu J_\nu+C_{\lambda \mu\rho} (x_\lambda +x_\mu)
Q_\lambda Q_\mu J_\rho+\nonumber\\&&2x_\mu C_{\lambda\mu\rho}
Q_\lambda J_\mu Q_\rho -2x_\rho C_{\rho\lambda\sigma} J_\rho
J_\lambda {J_\sigma}' +\frac{1}{3}C_{\mu\nu\lambda}(x_\mu-x_\nu)
J_\mu J_\nu {J_\lambda}'-2 x_\mu C_{\mu\rho\nu}J_\mu J_\rho
{J_\nu}' ].\nonumber\\&&\label{(II)}
\end{eqnarray}
where now $\alpha = 1,2,...,7$, while $\rho,\sigma=1,2,4$ and
$\lambda,\mu,\nu=3,5,6,7$.\par
The complete set of equations of motion for the $N=4$ $(II)$ case
occupies $13$ pages in LaTex. The given file is available upon
request. Just like the $N=3$ case and contrary to the $N=4$ $(I)$ case,
these equations of motion cannot be easily compactified since the
field labels $1\leftrightarrow 2$, $3$, $4$, $5$ and $6\leftrightarrow 7$
all play a different role.
\par
Let us conclude this section with a final comment. The two classes $(I)$ and $(II)$
of $N=4$ solutions are obviously inequivalent. For what concerns the first class
we can notice that by suitably choosing the parameters $x_\alpha$'s being given by
$x_1=x_2=-8$ $(x_3=16)$, the resulting generalized KdV system extends the integrable
$N=4$ KdV based on the ``minimal $N=4$ SCA". This leaves the possibility that the $N=4$
$(I)$ KdV, for the given values of the $x_\alpha$'s parameters and for some $x\neq0$,
could be an integrable system. We plan to address this issue in the future.

\section{Conclusions.}

In this paper we investigated the issue of large $N$
supersymmetric extensions of the KdV equation. The construction of
extended supersymmetrizations is important in connection with
integrable hierarchies since extended supersymmetric theories
provide the unification of otherwise unrelated bosonic or lower
supersymmetric hierarchies. The case mentioned throughout the
paper of the integrable $N=4$ KdV based on the $N=4$ SCA, which
encompasses both the inequivalent $a=-2$ and the $a=4$ $N=2$ KdVs,
is a nice example of that.\par From what concerns the applications
of supersymmetry many good reasons are found to investigate
extended supersymmetries. We refer to \cite{pt} for a detailed
discussion of various aspects. In this cited paper the
matrix-representations of the $N$ extended supersymmetries have
been classified (see also \cite{gr2}). Besides matrix
representations however, just in the case of the $N=8$
supersymmetry, a specific realization for it can be obtained via
the non-associative division algebra of the octonions (a detailed
discussion of this topic can be found in \cite{top}). From the
point of view of superextensions of KdV, it would be then quite
natural to expect the octonionic realization of the $N=8$
supersymmetry being related with the ``Non-associative $N=8$
Superconformal algebra" introduced in \cite{eng}. The
``non-associativity" is here referred to the fact that this
algebra does not satisfy the (super) Jacobi identities. This
apparent drawback turns out to be an advantage since it allows to
overcome a no-go theorem which prevented so far to construct
$N$-supersymmetric KdVs for $N>4$, due to the fact that no central
extension is allowed for superconformal algebras (of standard
type) for $N>4$ (see \cite{gls}).\par In the present paper we used
the ``Non-associative $N=8$ SCA" as a tool to produce the first
example of an $N=8$ supersymmetric extension of the KdV equation.
The system under consideration involves the $8$ bosonic and the
$8$ fermionic fields entering the $N=8$ SCA. We constructed the
$N=8$ superKdV equations by deriving them from a generalized
hamiltonian system admitting the ``Non-associative $N=8$ SCA" as
generalized Poisson brackets. To our knowledge this is also the
first example of a (generalized) dynamical system associated to
the given $N=8$ SCA.\par The main results of this paper can be
summarized as follows. We reviewed at first the $N=4$ KdV based on
the ``minimal $N=4$ SCA" and constructed the fundamental domain
for its inequivalent supersymmetrizations. Later we investigated
the possibility for an $N=8$ superKdV based on the $N=8$ SCA. We
arrived at a uniquely specified system of equations given by
formula (\ref{EOMN8}). This system corresponds to the $N=8$
superextension of the most symmetric (the $SU(2)$-invariant) point
in the fundamental domain of the $N=4$ KdV. Despite its enlarged
symmetry this point is however not an integrable point of the
$N=4$ KdV.\par In the following we investigated which $N$
supersymmetric extensions (for $N>2$) of KdV are supported by the
``Non-associative $N=8$ SCA" generalized Poisson brackets. The
complete results are stated as follows. Besides the unique $N=8$
case, such extensions are found for $N=3$ and $N=4$.\par The class
of solutions of the $N=3$ case depends on $6$ free parameters and
is reported in formula (\ref{hamn3}). For what concerns the $N=4$
cases two inequivalent classes of solutions, named ``$(I)$" and
``$(II)$", are found. The first class depends on three free
parameters, while the second one depends on just two free
parameters. They are given in formulas (\ref{(I)}) and
(\ref{(II)}) respectively. For a convenient choice of the
parameters of the class $(I)$ solution, the resulting system of
equations generalizes the integrable point of the ``minimal" $N=4$
KdV, leaving room to the possibility that a global $N=4$ system
involving the whole set of $N=8$ SCA fields could correspond to an
integrable hierarchy. This is an issue that we are planning to
address in a future work.

\end{document}